\documentclass[reprint,superscriptaddress,showpacs,amsmath,amssymb,aps,prl]{revtex4-1}
\usepackage{amsmath}
\usepackage{amssymb}
\usepackage{graphics}
\usepackage{graphicx}% Include figure files
\usepackage{dcolumn}% Align table columns on decimal point
\usepackage[pdfpagemode=UseNone,pdfstartview=FitH,colorlinks=true,linkcolor=blue,urlcolor=blue,anchorcolor=blue,citecolor=blue]{hyperref}
\allowdisplaybreaks[4]

\begin{document}

\title{Quantum Simulation of the Fermion-Boson Composite Quasi-Particles with a Driven Qubit-Magnon Hybrid Quantum System}

\author{Yi-Pu Wang}
\affiliation{Zhejiang Province Key Laboratory of Quantum Technology and Device, Department of Physics and State Key Laboratory of Modern Optical Instrumentation, Zhejiang University, Hangzhou 310027, China}
\affiliation{Quantum Physics and Quantum Information Division, Beijing Computational Science Research Center, Beijing 100193, China}

\author{Guo-Qiang Zhang}
\affiliation{Zhejiang Province Key Laboratory of Quantum Technology and Device, Department of Physics and State Key Laboratory of Modern Optical Instrumentation, Zhejiang University, Hangzhou 310027, China}
\affiliation{Quantum Physics and Quantum Information Division, Beijing Computational Science Research Center, Beijing 100193, China}

\author{Da Xu}
\affiliation{Zhejiang Province Key Laboratory of Quantum Technology and Device, Department of Physics and State Key Laboratory of Modern Optical Instrumentation, Zhejiang University, Hangzhou 310027, China}

\author{Tie-Fu  Li}
\thanks{litf@tsinghua.edu.cn}
\affiliation{Institute of Microelectronics, Tsinghua University, Beijing 100084, China}
\affiliation{Beijing Academy of Quantum Information Sciences, 100193 Beijing, China}

\author{Shi-Yao Zhu}
\affiliation{Zhejiang Province Key Laboratory of Quantum Technology and Device, Department of Physics and State Key Laboratory of Modern Optical Instrumentation, Zhejiang University, Hangzhou 310027, China}

\author{J. S. Tsai}
\affiliation{Department of Physic, Tokyo University of Science, Kagurazaka, Shinjuku-ku, Tokyo 162-8601, Japan}
\affiliation{Center for Emergent Matter Science, RIKEN, Wako-shi, Saitama 351-0198, Japan}

\author{J. Q. You}
\thanks{jqyou@zju.edu.cn}
\affiliation{Zhejiang Province Key Laboratory of Quantum Technology and Device, Department of Physics and State Key Laboratory of Modern Optical Instrumentation, Zhejiang University, Hangzhou 310027, China}

\date{\today}% It is always \today, today,
\begin{abstract}
We experimentally demonstrate strong coupling between the ferromagnetic magnons in a small yttrium-iron-garnet (YIG) sphere and the drive-field-induced dressed states of a superconducting qubit, which gives rise to the double dressing of the superconducting qubit. The YIG sphere and the superconducting qubit are embedded in a microwave cavity and the effective coupling between them is mediated by the virtual cavity photons. The theoretical results fit the experimental observations well in a wide region of the drive-field power resonantly applied to the superconducting qubit and reveal that the driven qubit-magnon hybrid quantum system can be harnessed to emulate a particle-hole-symmetric pair coupled to a bosonic mode. This hybrid quantum system offers a novel platform for quantum simulation of the composite quasi-particles consisting of fermions and bosons.
\end{abstract}

%\pacs{71.36.+c, 42.50.Pq, 76.50.+g, 75.30.Ds}

\maketitle

%\setcounter{secnumdepth}{3}
%\section{Introduction}
By exploiting the advantages of different components, hybrid quantum systems can provide outstanding architectures for quantum information processing (see, e.g., Refs. \cite{Xiang-13,Kurizki-15,Wallquist-09,Kimble-08}). Recently, a qubit-magnon hybrid quantum system was implemented by strongly coupling both a superconducting qubit and the Kittel mode of magnons in a small yttrium-iron-garnet (YIG) sphere to a three-dimensional (3D) microwave cavity~\cite{TabuchiScience-15,NakamuraSA-17}, where an effective appreciable qubit-magnon coupling was achieved by exchanging virtual cavity photons between the qubit and magnons. For the Kittel mode in the YIG sphere, the magnons are collective excitations of spins with zero wave number (i.e., in the long-wavelength limit) where all exchange-coupled spins in the sample precess uniformly~\cite{White-07}.
When hybridized with microwave cavity photons~\cite{SoykalPRL-10,SoykalPRB-10}, optical photons~\cite{haighpra-15}, and phonons~\cite{Tang-SA-16}, these magnons can provide a platform to implement various novel phenomena and applications~\cite{Huebl-13,Tabuchi-14,Zhang-14,Tobar-14,Hu-15,You-15,BauerPRB-15,CaoPRB-15,HaighPRB-15,LiuPRB-16,
Wangyp-16,SharmaPRB-17,SharmaPRL-18,OsadaPRL-18,Wu-18}, including magnon gradient memory~\cite{ZhangNC-15,Lambert-16}, cavity spintronics~\cite{Hu-15,HuPRL-17}, quantum transducer~\cite{Osada-16,Zhangxu-16,haighprl-16,Braggio-17}, parity-time symmetry~\cite{YouNC-17}, bistability of cavity magnon-polaritons~\cite{YouPRL-18,HuPRB-18}, and the generations of magnon-photon-phonon entanglement~\cite{Zhu-18} and squeezed magnon-phonon states~\cite{Zhu-19}. Moreover, ultrstrong-coupling regime between magnons and cavity photons have also be shown to be implementable~\cite{Tobar-14,TobarPRB-16,TobarAPL-16}, in addition to the strong couplings~\cite{Huebl-13,Tabuchi-14,Zhang-14,Tobar-14,Hu-15,You-15}.

The superconducting qubit as a quantum information processing unit can be used as the core of a solid-state hybrid quantum system~\cite{Xiang-13,Kurizki-15}. Coherent dressing of the qubit can provide access to a new quantum system with improved properties (e.g., longer coherence times~\cite{Laucht-17,Timoney-11}) and offer various applications in quantum information (see, e.g., Refs.~\cite{Meyer-13,Xue-17,Koshino-16}). Quantum mechanically, magnons behave similar to other bosons, so their strong coupling to the superconducting qubit can yield a coherent dressing of the qubit, as analogous to the dressing of a two-level system via photons~\cite{Scully}.

In this Letter, we study a driven qubit-magnon hybrid quantum system consisting of a 3D transmon qubit and a small YIG sphere embedded in a rectangular 3D microwave cavity. When driven in resonance by a monochromatic microwave field, the transmon qubit is dressed by the microwave field. Owing to the qubit-magnon coupling mediated by the virtual cavity photons, the 3D transmon qubit is further dressed by the magnons. From the dispersive readout results, we clearly show these newly-formed  doubly dressed states of the superconducting qubit by demonstrating their frequency splittings versus the drive power. The theoretical results fit the experimental observations well in a wide region of the drive power and reveal that this driven qubit-magnon hybrid system can be harnessed to emulate a particle-hole-symmetric pair coupled to a bosonic mode. These doubly dressed states behave like composite quasi-particles consisting of fermions and bosons and can be more fermion-like or not, depending on the portion of magnons in the hybridized normal modes. Our work opens a new avenue to quantum simulation of the fermion-boson composite quasi-particles with a hybrid quantum system.

\begin{figure}[tbp]
\centering
\includegraphics[width=0.4\textwidth]{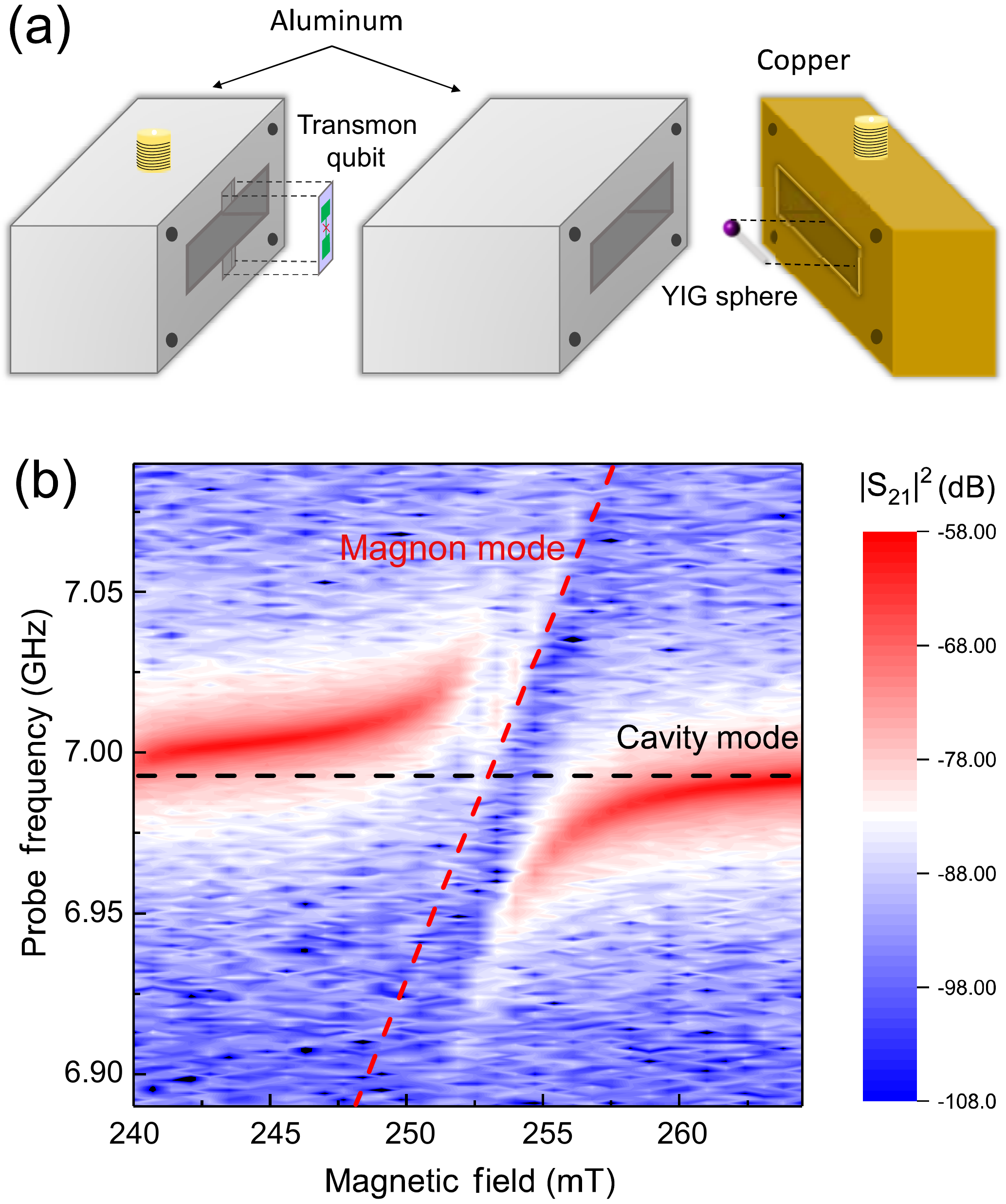}
\caption{(color online). (a) Schematic of the qubit-magnon hybrid quantum system in a rectangular 3D microwave cavity. A small YIG sphere is placed in the part of the cavity made of oxygen-free copper at the magnetic-field antinode of the cavity mode $\rm{TE}_{\rm{102}}$. To mount the transmon qubit, the part of the cavity made of aluminum is further separated into two parts that clamp the qubit at the electric-field antinode of the cavity mode $\rm{TE}_{\rm{102}}$. (b) Transmission spectrum of the cavity when the Kittel mode of magnons in the YIG sphere is magnetically tuned to be near resonance with the cavity mode $\rm{TE}_{\rm{102}}$.}
\label{fig:1}
\end{figure}

Figure~\ref{fig:1}(a) schematically show the experimental setup of the hybrid quantum system. The rectangular 3D microwave cavity has inner dimensions of $58\times 32\times 6$~mm$^3$. It consists of two parts which are made of aluminum and oxygen-free copper, respectively. The cavity is placed in a BlueFors LD-400 dilution refrigerator at a base cryogenic temperature of $\sim 20$~mK. At such a low temperature, the aluminum is superconducting and becomes diamagnetic. The 3D transmon qubit~\cite{Paik-2011,Jin-15} mounted in the superconducting aluminum part of the cavity can be protected from the magnetic field generated by the outside superconducting magnet. In this superconducting qubit, two aluminum pads are attached to the small Josephson junction, which, together with the cavity, provide a large shunt capacitor to suppress the charge noise in the qubit, as in the capacitively-shunted flux qubit~\cite{You-07,Yan-16}, 2D transmon~\cite{Koch-07}, and Xmon~\cite{Martinis-14}. A small YIG sphere of diameter 1 mm is placed in the copper part of the cavity which is not superconducting at the base temperature of the dilution refrigerator. The static magnetic field produced by the outside superconducting magnet can go into the copper part of the cavity to adjust the frequency of the Kittel mode in the YIG sphere. The static magnetic field is aligned along the hard magnetization axis [100] of the YIG sphere and both the 3D transmon qubit and the Kittel mode are strongly coupled to the cavity mode $\rm{TE}_{\rm{102}}$. The frequency of this cavity mode $\rm{TE}_{\rm{102}}$, which is measured to be $\omega_c/2\pi\approx 6.99$~GHz, is designed to have a large detuning from the transition frequency of the 3D transmon qubit. When the Kittel mode is tuned to be nearly resonant with the qubit by the static magnetic field (i.e., the cavity mode $\rm{TE}_{\rm{102}}$ is also largely detuned from the Kittel mode), an effective qubit-magnon coupling can be achieved via the exchange of virtual cavity photons between the qubit and magnons~\cite{TabuchiScience-15,NakamuraSA-17}.

Before exhibiting the coupling between the 3D transmon qubit and the Kittel mode of magnons, we first measure the coupling between the Kittel mode and the cavity mode $\rm{TE}_{\rm{102}}$ by tuning the frequency of the Kittel mode in resonance with this cavity mode. Owing to the strong coupling between them, two branch of magnon polaritons occur around the anticrossing point [Fig.~1(b)], with a level splitting $\sim 86$ MHz at the anticrossing point. The transition frequency between ground state $|g\rangle$ and first excited state $|e\rangle$ of the 3D transmon qubit is measured to be $\omega_{q}/2\pi\approx6.49$ GHz. Because the qubit is now largely detuned from both the Kittel mode and the cavity mode $\rm{TE}_{\rm{102}}$, its effect on the magnon polaritons can be neglected at around the anticrossing point. Therefore, from the measured level splitting at the anticrossing point, the coupling strength between the Kittel mode and the cavity mode $\rm{TE}_{\rm{102}}$ can be obtained as $g_{m}\approx 43.0$ MHz.

To achieve an effective coupling between the qubit and the Kittel mode, we then tune the frequency $\omega_{m}$ of the Kittel mode to be near the transition frequency $\omega_{q}$ of the qubit. Now, the cavity mode $\rm{TE}_{\rm{102}}$ is largely detuned from both the qubit and the Kittel mode, and an effective qubit-magnon coupling can be achieved via the virtual photons of the cavity mode $\rm{TE}_{\rm{102}}$~\cite{TabuchiScience-15,NakamuraSA-17}. At the cryogenic temperature $T\sim 20$~mK, $k_BT\ll \omega_{q}, \omega_{m}$, so both the qubit and the Kittel mode stay nearly in their ground states  $|g\rangle$ and $|0\rangle$, respectively, and a transition from $|e\rangle$ to $|g\rangle$ induces a transition from the vacuum to single-magnon states (i.e., from $|0\rangle$ to $|1\rangle$). As in Ref.~\cite{TabuchiScience-15}, the spectroscopic measurement is carried out with a vector network analyzer (VNA) by probing the transmission of the cavity mode $\rm{TE}_{\rm{103}}$. During measurement, the probe field is applied at the resonant frequency of the cavity mode $\rm{TE}_{\rm{103}}$.
%{\color{red} ($\sim *.**$ GHz)}.
Largely detuned from the cavity mode, the qubit is usually in the ground state. When the qubit is excited, the cavity mode has an observable frequency shift in this dispersive regime and the probed field transmission thereby changes. Also, a series of attenuators and isolators are used to prevent thermal noise from reaching the sample and the signal going out from the output port of the cavity is amplified by two low-noise amplifiers at the stages of 4K and room temperature, respectively. Moreover, two microwave fields from different sources are applied to the input port of the cavity. One is used as an excitation field and tuned to excite the hybridized normal modes of the system, and the other has a fixed frequency near or in resonance with the 3D transmon qubit and is harnessed to drive the qubit via the large capacitor shunted to the Josephson junction.

\begin{figure}[tbp]
\centering
\includegraphics[width=0.35\textwidth]{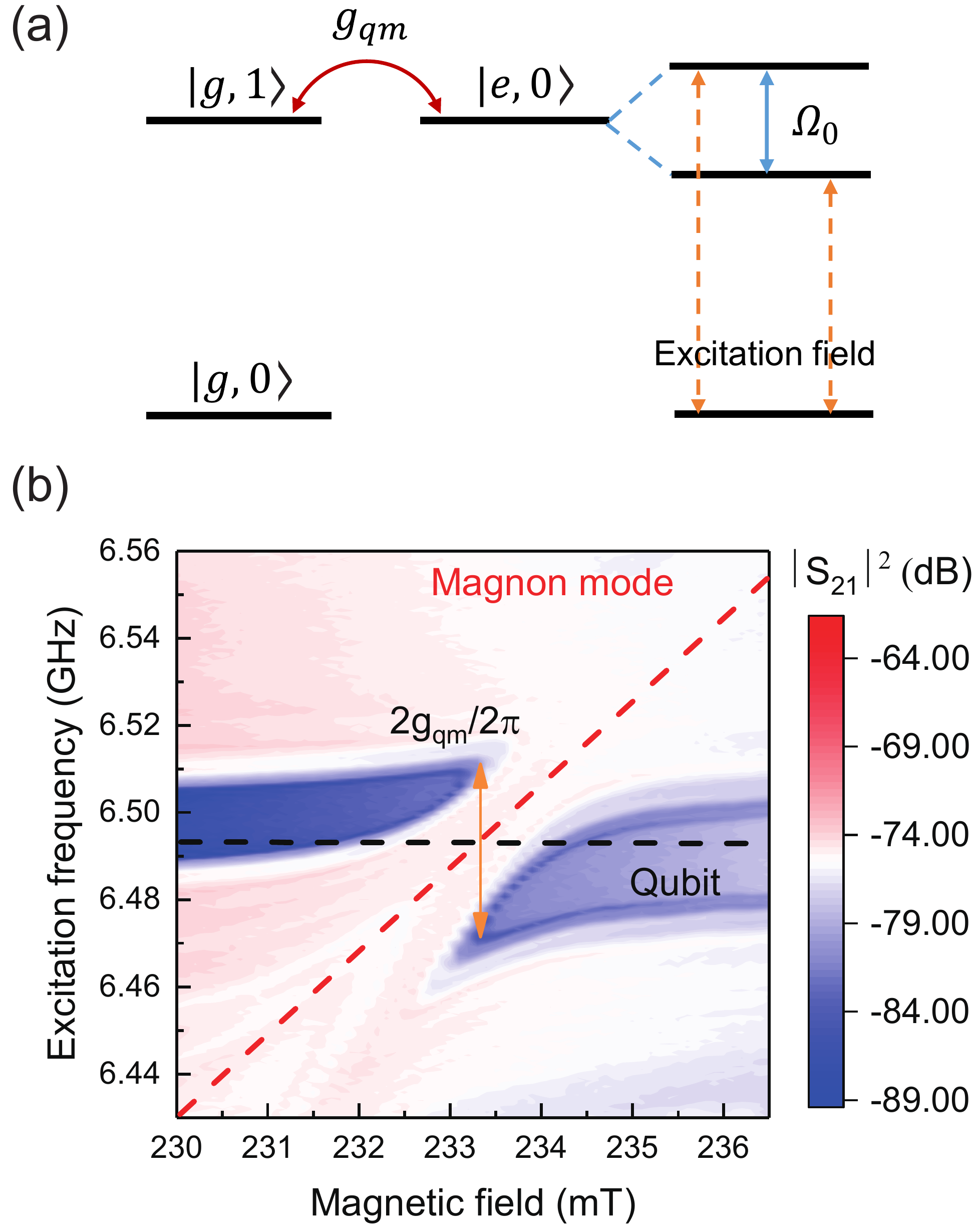}
\caption{(color online) (a) Energy levels of the qubit-magnon system with only the vacuum and single-magnon states involved for the Kittel mode. The coupling between $|g,1\rangle$ and $|e,0\rangle$ induces the vacuum Rabi splitting $\Omega_0$. (b) The vacuum Rabi splitting of the qubit-magnon system measured via the transmission spectrum of the cavity by tuning both the excitation field and the static magnetic field. The probe field is applied in resonance with the cavity mode $\rm{TE}_{\rm{103}}$.}
\label{fig:2}
\end{figure}

In Fig.~\ref{fig:2}(a), we show the energy levels of the qubit-magon hybrid quantum system at the cryogenic temperature when the drive field on the qubit is turned off. The interaction between $|g,1\rangle$ and $|e,0\rangle$ gives rise to the vacuum Rabi splitting $\Omega_0$. Owing to the exchange of virtual cavity photons between the qubit and the Kittel mode, the effective qubit-magnon coupling is~\cite{SM} $g_{qm}=\frac{1}{2}g_qg_m(1/\Delta_q+1/\Delta_m)$, where $\Delta_q\equiv\omega_c-\omega_q$ ($\Delta_m\equiv\omega_c-\omega_m$) is the frequency detuning of the cavity mode $\rm{TE}_{\rm{102}}$ from the qubit (Kittel mode). The effective Hamiltonian of the coupled qubit-magnon system is given by (we set $\hbar=1$)
\begin{equation}
H_{qm}=\frac{\omega_q}{2}\sigma_z+\omega_mb^{\dag}b+g_{qm}(\sigma_{+}b+\sigma_{-}b^{\dag}),
\end{equation}
where $\sigma_z$ and $\sigma_{\pm}$ are Pauli operators of the qubit and $b^{\dag}$ ($b$) is the creation (annihilation) operator of the magnons.
When the Kittel mode is tuned to be in resonance with the 3D transmon qubit, $g_{qm}$ is reduced to~\cite{TabuchiScience-15} $g_{qm}=g_qg_m/\Delta$, with $\Delta_q=\Delta_m=\Delta$. The coupling between the qubit and the Kittel mode is measured in Fig.~\ref{fig:2}(b) by tuning the static magnetic field and scanning the frequency of the excitation field. From the vacuum Rabi splitting measured at the anticrossing point ($\Omega_0=2g_{qm}$), we obtain $g_{qm}/2\pi\approx20.1$ MHz.
%which is larger than the achieved coupling strength $g_{qm}/2\pi\approx10.0$ MHz in Ref.~\cite{TabuchiScience-15}.
Comparing the linewidth of the mode at around 230 and 237 mT, we see that the linewidth becomes broader at a stronger magnetic field, indicating that some static magnetic field penetrates into the cavity to affect the quantum coherence of the qubit.
Actually, in our experiment, the static magnetic field was already designed to be parallel to the chip surface of the qubit, so as to reduce its influence on the qubit as much as possible.
With the measured $\Delta\approx 0.5$ GHz, $g_m\approx 42.0$ MHz and $g_{qm}/2\pi\approx20.1$ MHz, we can deduce using $g_{qm}=g_qg_m/\Delta$ that $g_q\approx 239$ MHz, which reveals that the interaction between the qubit and the cavity mode $\rm{TE}_{\rm{102}}$ is in the strong-coupling regime.

When turning on the monochromatic drive field of frequency $\omega_d$ appied to the 3D transmon qubit, the Hamiltonian of the driven qubit-magnon hybrid system is written as $H_d=H_{qm}+\frac{1}{4}\Omega_d(\sigma_{+}e^{-i\omega_d t}+\sigma_{-}e^{i\omega_d t})$, where $\Omega_d$ represents the coupling strength between the drive field and the qubit.
In the rotating frame with respect to the drive frequency $\omega_d$, the Hamiltonian of this driven qubit-magnon system becomes
%\begin{equation}
$H=\frac{1}{2}\delta_q\sigma_z+\frac{1}{2}\Omega_d\sigma_x+\delta_mb^{\dag}b+g_{qm}(\sigma_{+}b+\sigma_{-}b^{\dag})$,
%\end{equation}
where $\delta_q\equiv\omega_q-\omega_d$ ($\delta_m\equiv\omega_m-\omega_d$) is the frequency detuning between the qubit (magnon) and the drive field. The qubit is now dressed by a classical drive field and the corresponding Rabi frequency is $\widetilde{\Omega}_d=\sqrt{\Omega_d^2+\delta_q^2}$. Due to the coupling between the qubit and magnons, the qubit is then further dressed by the magnons. Therefore, the driven qubit-magnon hybrid system can be described as a {\it doubly dressed qubit}.
Below we focus on the resonant case with $\delta_q=\delta_m=0$. The Hamiltonian $H$ becomes
\begin{equation}
H=\frac{\Omega_d}{2}\sigma_x+g_{qm}(\sigma_{+}b+\sigma_{-}b^{\dag}).
\label{eq2}
\end{equation}
Even for this simple model, one cannot exactly solve it, because the model of a double dressed qubit is not exactly solvable~\cite{Shevchenko-14}.
With the new basis states $|\pm\rangle=(|g\rangle\pm|e\rangle)/\sqrt{2}$ and the mean-field approximation, the Hamiltonian (\ref{eq2}) can be reduced to~\cite{SM} $H=H_p+H_h$, with
\begin{eqnarray}
H_p&=&\frac{\Omega_d}{2}a^{\dag}a+\tilde{g}_{qm}(a^{\dag}b+ab^{\dag}),\nonumber\\
H_h&=&-\frac{\Omega_d}{2}h^{\dag}h-\tilde{g}_{qm}(h^{\dag}b+hb^{\dag}),
\end{eqnarray}
where $a=h^{\dag}\equiv|-\rangle\langle+|$, and $\tilde{g}_{qm}=\frac{1}{2}g_{qm}(A+1)$, with $A=\langle a\rangle$. Because $\{a,a^{\dag}\}=1$ and $\{a,a\}=\{a^{\dag},a^{\dag}\}=0$, these newly-defined operators are fermionic.
Here $H_h$ has the same form as $H_p$, but its two parameters $\Omega_d$ and $\tilde{g}_{qm}$ change the signs. This means that the Hamiltonian $H=H_p+H_h$ has the {\it particle-hole symmetry}. Thus, we can use the driven qubit-magnon hybrid system to emulate a particle-hole-symmetric pair of fermions coupled to a bosonic mode.

\begin{figure}
\centering
 \includegraphics[width=0.49\textwidth]{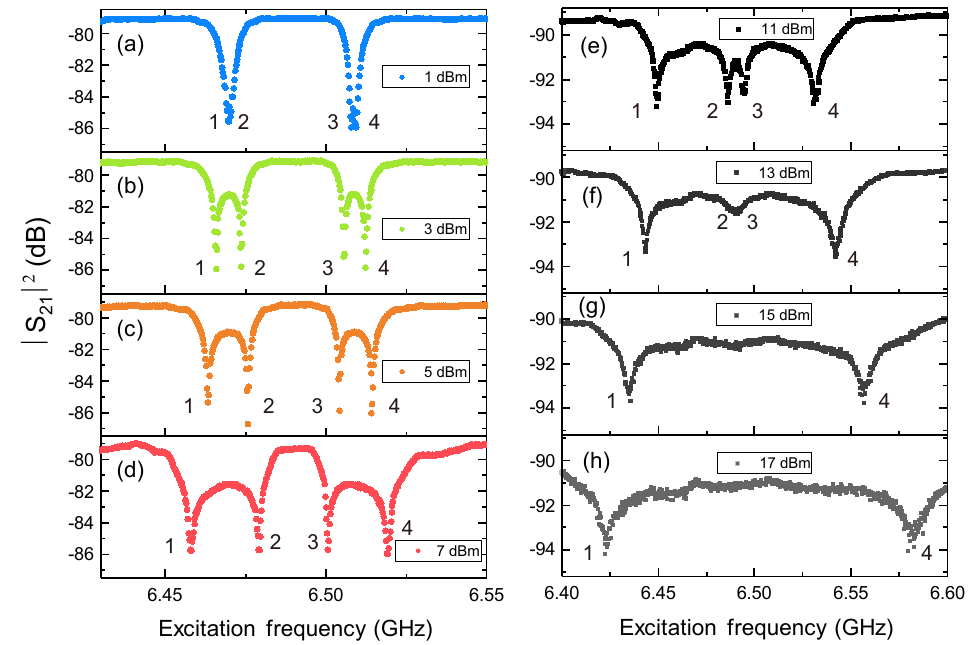}
 \caption{(color online) Dispersive readout of the hybridized normal modes of the driven qubit-magnon system. An excitation field is tuned to excite the hybridized normal modes and a probe field is applied in resonance with the cavity mode $\rm{TE}_{\rm{103}}$. The power of the microwave field to drive the superconducting qubit is tuned to be (a) 1 dBm, (b) 3 dBm, (c) 5 dBm, (d) 7 dBm, (e) 11 dBm, (f) 13 dBm, (g) 15 dBm, and (h) 17 dBm, respectively.}
 \label{fig3}
\end{figure}

The two eigenvalues of the Hamiltonian $H_p$ (denoted as $\omega_4$ and $\omega_2$) are $\lambda_{\pm}=\frac{1}{2}[(\Omega_d/2)\pm[(\Omega_d/2)^2+(2\tilde{g}_{qm}\sqrt{N+1})^2]^{1/2}$, where $N$ is the number of magnons excited. The eigenstates correspond to the particle-magnon composite quasi-particle states. The two eigenvalues of $H_h$ (denoted as $\omega_1$ and $\omega_3$) are ${\overline\lambda}_{\pm}=-\lambda_{\pm}$, and the eigenstates correspond to the hole-magnon composite quasi-particle states.  By tuning $\Omega_d$, these hybridized normal modes can be more fermion-like or not, depending on the portion
of magnons in these modes. The frequency splitting between modes 4 and 1 and the frequency splitting between modes 3 and 2 are
\begin{eqnarray}
\omega_4-\omega_1&=&\frac{\Omega_d}{2}+\left[\left(\Omega_d/2\right)^2+(2\tilde{g}_{qm}\sqrt{N+1})^2\right]^{1/2},\nonumber\\
\omega_3-\omega_2&=&-\frac{\Omega_d}{2}+\left[\left(\Omega_d/2\right)^2+(2\tilde{g}_{qm}\sqrt{N+1})^2\right]^{1/2}.~~~~
\label{eq4}
\end{eqnarray}
In Fig.~\ref{fig3}, we display the transmission spectrum of the driven hybrid system measured in the resonant case of $\delta_q=\delta_m=0$ by successive increasing the power $P_d$ of the drive field. The spectrum exhibits a clear mirror symmetry owing to the particle-hole symmetry in the Hamiltonian. When increasing $P_d$, two absorption dips are split into four, with the outer two dips (labelled as 1 and 4) departing away. Accordingly, the inner two dips (labelled as 2 and 3) gradually approach and then merge together. The Rabi frequency $\Omega_{d}$ is proportional to the drive-field amplitude $\varepsilon$, so $\Omega_{d}\propto\sqrt{P_{d}}$. From Eq.~(\ref{eq4}) it follows that $\omega_4-\omega_1$ increases and $\omega_3-\omega_2$ decreases when increasing $P_{d}$. In particular, for a sufficiently strong drive field with $\Omega_d\gg  4\tilde{g}_{qm}\sqrt{N+1}$, $\omega_3-\omega_2\rightarrow 0$. These behaviors agree with the observations in Fig.~\ref{fig3}.

\begin{figure}
\centering
 \includegraphics[width=0.48\textwidth]{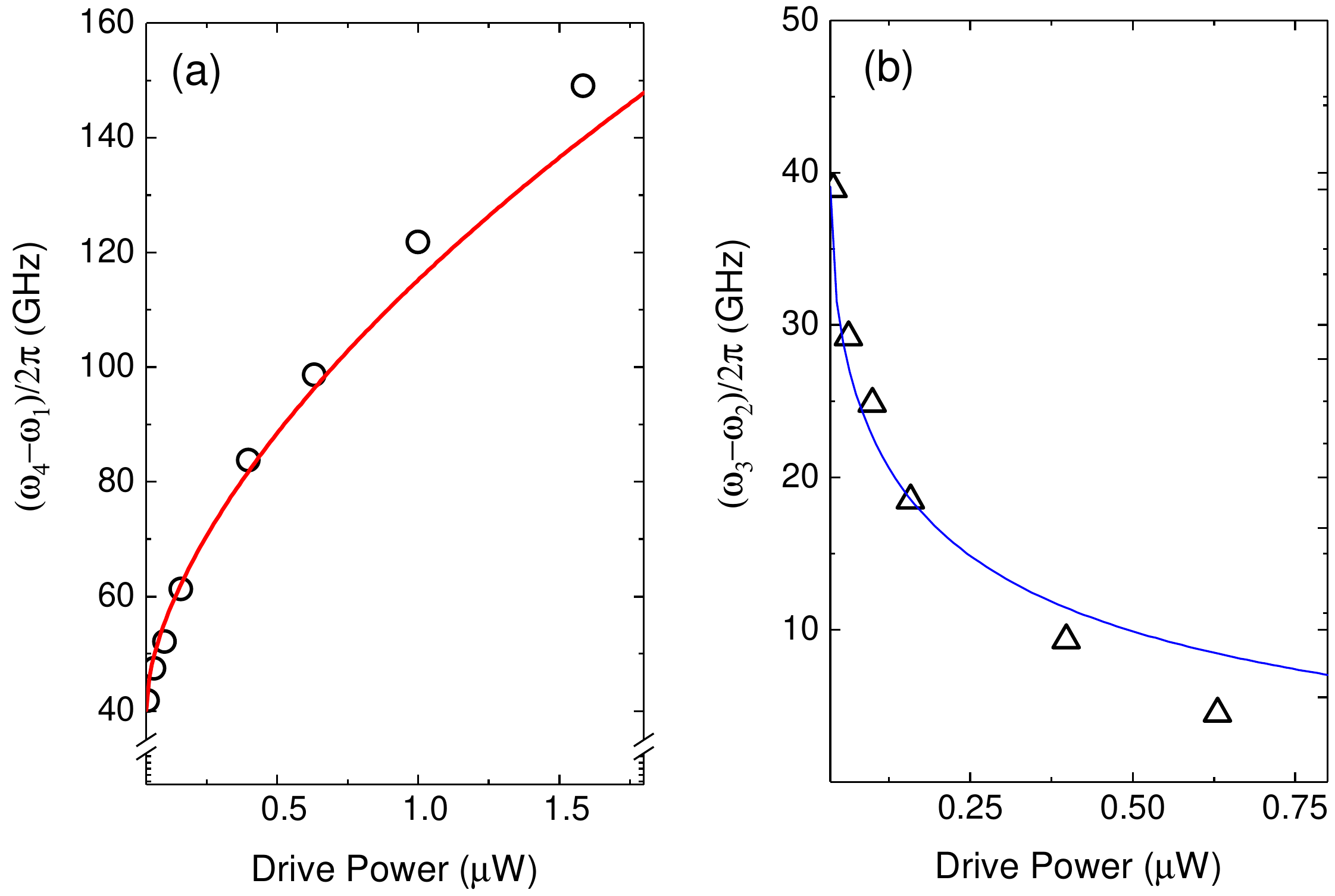}
 \caption{(color online) (a) Fitting the experimental data of the frequency splitting between hybridized normal modes 1 and 4 versus the drive power $P_{d}$. (b) Fitting the experimental data of the frequency splitting between hybridized normal modes 2 and 3 versus the drive power $P_{d}$. To fit the data, we use Eq.~(\ref{eq4}), where $\tilde{g}_{qm}=g_{qm}$, $N=0$, and $\Omega_{d}=k\sqrt{P_{d}}$, with $k=103$~MHz/${\rm mW}^{1/2}$.}
 \label{fig4}
\end{figure}

For zero $\Omega_d$ (i.e., when turning off the drive field), Eq.~(\ref{eq4}) gives $\omega_4-\omega_1=\omega_3-\omega_2=2\tilde{g}_{qm}\sqrt{N+1}$. In fact, when $\Omega_d=0$, the model in Eq.~(\ref{eq2}) becomes exactly solvable, with the Rabi splitting $\Omega_N=2g_{qm}\sqrt{N+1}$. In this limiting case, we obtain $\tilde{g}_{qm}\equiv \frac{1}{2}g_{qm}(A+1)=g_{qm}$, i.e., $A=1$. In Fig.~\ref{fig4}, we display the frequency splittings $\omega_4-\omega_1$ and $\omega_3-\omega_2$ versus the drive power $\Omega_d$, with the data extracted from Fig.~\ref{fig3}.
Fitting these data, we use Eq.~(\ref{eq4}), where $\Omega_{d}=k\sqrt{P_{d}}$ and $\tilde{g}_{qm}$ is replaced by $g_{qm}$. With $N=0$, we obtain the only fitting parameter $k=103$~MHz/${\rm mW}^{1/2}$. Here the attenuation of the drive field from the microwave source to the input port of the cavity is 45 dB. In the region of a weaker drive field, the numerical results with $N=0$ agree well with the experimental results. This reveals that for a weaker drive field, the Kittel mode still stays nearly in the ground state and the drive field does not appreciably affect it. However, when the drive field becomes strong, the numerical results with $N=0$ deviate from the experimental observations, i.e., smaller (larger) than the data for $\omega_4-\omega_1$ ($\omega_3-\omega_2$). This is what we expect because more magnons may be excited when the drive field is strengthened to raise the temperature in the cavity.

In conclusion, we have convincingly demonstrated strong coupling between the Kittel-mode magnons in a small YIG sphere and the drive-field-induced dressed states of a 3D transmon qubit. The coupling between the magnons and the dressed states of the superconducting qubit is mediated by the virtual cavity photons. The theoretical results fit the experimental observations well in a wide region of the drive power of the microwave field resonantly applied to the transmon qubit.
Our work opens up new possibilities to manipulate a hybrid quantum system composed of several degrees of freedom and paves the way to quantum simulation of composite quasi-particles consisting of fermions and bosons.

\begin{acknowledgments}
This work was supported by the National Key Research and Development Program of China (Grant No.~2016YFA0301200) and the National Natural Science Foundation of China (Grants No.~U1801661 and No. 11774022). T.-F.L. was partially supported by Science Challenge Project (Grant No. TZ2018003). J.S. Tsai was supported by the JST CREST, NEDO IoT, and Japanese cabinet office ImPACT.

\end{acknowledgments}

\end{document}